\RequirePackage{fix-cm}
\documentclass[smallextended]{svjour3}       
\smartqed  
\usepackage{epsf,graphics,graphicx}
\newcommand \beq{\begin{eqnarray}}
\newcommand \eeq{\end{eqnarray}}
\def\simge{\mathrel{%
       \rlap{\raise 0.511ex \hbox{$>$}}{\lower 0.511ex \hbox{$\sim$}}}}
\def\simle{\mathrel{
       \rlap{\raise 0.511ex \hbox{$<$}}{\lower 0.511ex \hbox{$\sim$}}}}

\newcommand \la{\raisebox{-.5ex}{$\stackrel{<}{\sim}$}}
\newcommand \ga{\raisebox{-.5ex}{$\stackrel{>}{\sim}$}}
\begin{document}

\title{Low Temperature Transport Properties of Very Dilute
Classical Solutions of $^3$He in Superfluid $^4$He
\thanks{This research was supported in part by NSF Grants PHY08-55569, PHY09-69790, and  PHY13-05891.   Author GB is grateful to the Aspen Center for Physics, supported in part by NSF Grant PHY-1066292, and the Niels Bohr International Academy, where parts of this research were carried out.}}
\titlerunning{Transport in $^3$He in Superfluid $^4$He}        

\author{Gordon Baym   \and
        D.~H.~Beck \and \nolinebreak C.~J.~Pethick
}

\authorrunning{G. Baym, D.H. Beck, \&  C.J. Pethick} 

\institute{
Gordon Baym \at
Department of Physics, University of Illinois, 1110 W. Green Street, Urbana, IL 61801, USA, and \\
The Niels Bohr International Academy, The Niels Bohr Institute, University of Copenhagen, Blegdamsvej 17, DK-2100 Copenhagen \O, Denmark\\
\email{gbaym@illinois.edu}           
\and
D.\ H.\ Beck \at
Department of Physics, University of Illinois, 1110 W. Green Street, Urbana, IL 61801, USA
\and
C.\ J.\  Pethick \at
Department of Physics, University of Illinois, 1110 W. Green Street, Urbana, IL 61801, USA, \\
The Niels Bohr International Academy, The Niels Bohr Institute,  University of Copenhagen, Blegdamsvej 17, DK-2100 Copenhagen \O, Denmark, and \\
NORDITA, Roslagstullsbacken 23, SE-10691 Stockholm, Sweden
}

\date{Received: date / Accepted: date}

\maketitle

\begin{abstract}
  We report microscopic calculations of the thermal conductivity, diffusion constant and thermal diffusion constant for 
classical solutions of $^3$He in superfluid $^4$He at temperatures $T \la 0.6$~K, where phonons are the dominant excitations of the $^4$He.  We focus on solutions with $^3$He concentrations $\la \,10^{-3}$, for which the main scattering mechanisms are phonon-phonon scattering via 3-phonon Landau and Beliaev processes, which maintain the phonons in a drifting equilibrium distribution, and the slower process of $^3$He-phonon scattering, which is crucial for determining the  $^3$He distribution function in transport.   We use  the fact that the relative changes in the energy and momentum of a $^3$He atom in a collision with a phonon are small to derive a Fokker-Planck equation for the $^3$He distribution function, which we show has an analytical solution in terms of Sonine polynomials.  We also calculate the corrections to the Fokker-Planck results for the transport coefficients. 
\end{abstract}

\paragraph{\noindent\date{\today}}

\section{Introduction}

\begin{center}{\it Problems worthy of attack, prove their worth by hitting back} \cite{grook}\\  \end{center}

Historically, the study of equilibrium and transport processes in the helium liquids revealed important information about the basic physics of quantum liquids \cite{reviews}.  A proposed experiment at Oak Ridge National Laboratory \cite{snsExpt} to search sensitively for the electric dipole moment of the neutron has renewed interest in the transport properties of dilute solutions of $^3$He in liquid $^4$He in low concentration regimes never before studied in detail.   The experiment will look for the effect of an electric field on the capture rate of polarized neutrons on polarized $^3$He atoms dissolved in $^4$He. 

  In a relative spin singlet state, the  capture rate of a neutron on a $^3$He atom can proceed through a virtual $\alpha$ particle state, and is thus enormous compared with the rate for a relative triplet state.   The experiment therefore aims to detect the precession of the neutron in an electric field from a change in the capture rate.  With time, however, the $^3$He atoms depolarize through scattering against the container walls,  and in the course of the experiment they will be driven out of the experimental volume by a phonon wind generated by a heater in the liquid $^4$He, and replaced with fresh polarized $^3$He.
  
    A novel pilot experiment was performed by Lamoreaux et al.
\cite{lamoreaux}, who measured the $^3$He density gradient induced by a
heat flow at temperatures $\la$ 0.6 K, at which the effects of rotons
are negligible.  In this experiment, the $^3$He number concentration,
$x_3 = n_3/(n_3+n_4)$, where $n_3$ and $n_4$ are the $^3$He and $^4$He
number densities, lay in the range $ 7\times 10^{-5}$ to $1.5\times
10^{-3}$ in the non-degenerate regime. The experiment was initially
interpreted in terms of diffusion of $^3$He atoms with respect to a gas
of phonons \cite{lamoreaux,bowley}, but in Ref.~\cite{highx} we showed
that the relevant transport coefficient is the total thermal
conductivity of the liquid, which consists of two contributions, one
from the $^3$He and another from diffusive flow of phonons relative to $^3$He.
This latter diffusion does not
necessarily involve net relative motion of $^4$He atoms relative to
$^3$He, as does diffusion in a normal system, since the net flux of $^4$He
atoms has a contribution from the superfluid as well as that from the
phonons. 

    In transport in this relatively high-$x_3$ regime, the $^3$He-$^3$He interactions are sufficiently strong that they keep the $^3$He in thermal equilibrium at rest at the local temperature $T(\vec r\,)$, while  phonon-phonon interactions keep the phonons in drifting local equilibrium.   In thermal transport, phonons transfer momentum to the  $^3$He atoms, via the $^3$He-phonon interactions.   The calculations in Ref.\ \cite{highx} took into account a number of physical effects not included in earlier calculations of the thermal conductivity.  Among these are phonon-phonon scattering, which rapidly establishes thermal equilibrium among phonons traveling in a given direction, $^3$He-$^3$He scattering, which is effective in maintaining thermal equilibrium of these atoms, and energy transfer in $^3$He-phonon collisions, which reduces the rate of these collisions.  These calculations agree well with the data of Ref.~\cite{lamoreaux}. 
    
    By contrast, in the proposed search for the neutron electric dipole moment,  the concentration of $^3$He will be in the much lower range $10^{-9}$-$10^{-11}$.   Here thermal conduction is essentially limited by transfer of phonon momentum to the container walls by viscous forces, with the $^3$He playing a negligible role.   Furthermore, collisions of $^3$He with the phonons drive the $^3$He away from equilibrium. 
    
    In this paper, starting from microscopic scattering processes we calculate transport coefficients of dilute solutions in the regime where phonons dominate the $^4$He excitations and the $^3$He concentrations are low enough for classical statistics to be valid.  This regime encompasses the range at relatively high $^3$He concentrations where measurements have been carried out~\cite{lamoreaux,rosenbaum}, to the range of natural concentration, $x_3 \sim 10^{-6}$ where future measurements are being prepared~\cite{harvard},  to the low concentrations anticipated in the Oak Ridge neutron electric dipole moment experiment~\cite{snsExpt}.  Our primary focus here is on the processes of thermal conduction and diffusion.

In the high-$x_3$ regime, $^3$He-$^3$He collisions maintain the $^3$He
quasiparticles in a drifting equilibrium distribution.   For this case
as we showed in Ref.~\cite{highx}, in calculating the leading
low-temperature contributions to the transport coefficients, one may
neglect the energy transfer in  $^3$He-phonon collisions. At low $x_3$, however,
the effects of $^3$He-$^3$He collisions are negligible and  as we show, 
it is necessary to take into account the energy transfer even in
calculating the leading low-temperature properties.  The reason for this
is that while the fractional changes in the momentum and energy of a $^3$He
quasiparticle in scattering by a phonon are both small, they are of the same
order, $\sim k/p\sim (T/m^* s^2)^{1/2}$, where $\bf k$ is the momentum
transfer, $p$ the momentum of a quasiparticle, $m^*$ the $^3$He effective mass, and $s$ the sound
velocity in $^4$He.
To calculate the leading low-temperature behavior we expand the
collision term in the $^3$He Boltzmann equation to leading non-trivial
(second) order in $k/p$ and derive a Fokker-Planck equation for
the distribution function.   Quite remarkably, we find that the
eigenfunctions of the collision integral may be found analytically and
have the form of Sonine polynomials, which are familiar in the theory of
transport coefficients in classical, single-component gases \cite{LP,sonine}.

   We begin in Sec.~\ref{sec:conservationLaws} by reviewing the conservation laws obeyed by the phonons and $^3$He, and define the particle and heat currents in detail. This section follows in large part the general approach to the dissipative hydrodynamics of mixtures in Ref.~\cite{LLFM}, but with crucial differences.  Then in Sec.~\ref{sec:scattering}, we review the microscopic $^3$He-$^3$He interactions, $^3$He-phonon interactions, and phonon-phonon interactions underlying the transport.  Here we do not consider effects, such as phonon scattering from the walls, whose role depends on the specific geometry in question. The effects of walls, which are critical at ultralow $x_3$ as will be encountered in the nEDM experiment, will be discussed in Ref.~\cite{BBPlowx}.
The scattering of phonons on $^3$He is characteristically that of a massless particle, as in Rayleigh scattering.  We then develop the $^3$He and phonon Boltzmann equations in Secs.~\ref{sec:3HeB} 
and \ref{sec:phononB}, and discuss recoil corrections in Sec. \ref{sec:recoil}. The resulting transport coefficients are presented in Sec.~\ref{sec:results}.  In general we work in the linear approximation, assuming that the deviations from equilibrium are small.

\section{Hydrodynamics and Conservation Laws}
\label{sec:conservationLaws}

  In dilute solutions of $^3$He in low temperature $^4$He in bulk, mean free paths are generally sufficiently small that the system can be described in terms of  dissipative hydrodynamics.   In addition, at low temperatures the dominant bosonic excitations of solutions are phonons.  We begin by laying out the general framework of the hydrodynamics, explicitly in terms of the $^3$He and $^4$He number densities,  the superfluid velocity, and the normal fluid $^3$He and phonon velocities.  In a later section we calculate the needed transport coefficients.   The description we give here parallels in many ways the description of a normal two-component fluid in Ref.~\cite{LLFM}, as extended to dilute solutions by Khalatnikov and Zharkov~\cite{KandZ,khalat}, with the identification of the phonons and the $^3$He as the two fluids.  An important difference from two component normal fluids is that here the excitations of the $^4$He are not conserved, and furthermore a phonon drift with respect to the $^3$He is a heat flow, whereas in a two-component normal fluid, heat flow is energy transport in the absence of relative motion of the two components. 
    
   \subsection{Conservation of particle number and momentum}
 
  The $^3$He excitation spectrum is effectively
\beq
  \epsilon_p = \epsilon_0 + p^2/2m^*,
\eeq  
where the $^3$He effective mass  is 
 $m^*=m_3 +\delta m \simeq 2.34 m_3$ \cite{BE67}; the $\delta m$ arises from backflow in the $^4$He as a $^3$He moves through it.    The mean field shift $\epsilon_0$ depends on both the $^3$He and $^4$He densities; however at low concentrations the former dependence is negligible, and $\partial \epsilon_0/\partial n_4 = (m_4s^2/n_4)(1+\alpha)$ where $\alpha \simeq 0.28$ is the fractional excess molar volume of the $^3$He.
At the temperatures and concentrations of interest, the $^3$He atoms obey Boltzmann statistics, with the equilibrium density given by
\beq
  n_3 = \nu\int\frac{d^3p}{(2\pi)^3}e^{-\beta (p^2/2m^* +\epsilon_0 - \mu_3)} =\nu e^{\beta(\mu_3-\epsilon_0)}/\lambda_{th}^3,
  \label{n3a}
\eeq
where $\mu_3$ is the $^3$He chemical potential, $\nu$ is the number of spin degrees of freedom: $\nu=1$ in a fully spin-polarized gas and 2 in an unpolarized gas, $\beta = 1/T$,
and $\lambda_{th} =  \sqrt{2\pi/m^*T}$ is the $^3$He thermal wavelength. 
We generally
work in units with $\hbar = k_B = 1$.   At the low concentrations of interest, effects of $^3$He-$^3$He interactions on the  thermodynamics are negligible.
 
   The conservation law for $^3$He atoms is 
\beq
 \frac{\partial n_3}{\partial t} + \vec\nabla\cdot(n_3\vec v_3)=0,
 \label{3cons}
\eeq
while the conservation law for $^4$He is
\beq
m_4 \frac{\partial n_4}{\partial t} + \vec\nabla\cdot \vec g_4=0,
\label{4cons}
\eeq
where $\vec v_3$ is the $^3$He flow velocity, and $\vec g_4 = \rho_s \vec v_s + \delta m \,n_3 \vec v_3 + \rho_{ph}\vec v_{ph}$ is the $^4$He mass current, with  $\vec v_s$ the superfluid velocity,  $
\vec v_{ph}$ the phonon (normal) fluid velocity, and $m_4$ the $^4$He atomic mass.   The $^4$He mass density is $\rho_4 = m_4 n_4$.   The 
superfluid mass density is $\rho_s =  \rho_4 - \rho_{ph} - \delta m\, n_3$, since the $^3$He effective mass correction
does not participate in superfluid flow.  The
 phonon fluid velocity is defined by writing the momentum density carried by phonons as $\rho_{ph}\vec v_{ph}$, with the normal mass density of the phonons given by
\beq
\rho_{ph} = \beta \int\frac{d^3q}{(2\pi)^3} \frac{q^2}{3}n_q^0(1+n_q^0) =\frac{2\pi^2}{45}\frac{T^4}{s^5},
\label{rhonq}
\eeq
where $s$ is the first sound velocity, $\vec q$ the phonon momentum, and $n_q^0 = \left(e^{\beta s q} -1\right)^{-1}$.   

    The mass current, or total momentum density, $\vec g$, in the solution is given by
\beq
\vec g = \rho_s\vec v_s + \rho_{ph}\vec v_{ph} + m^*\,n_3\vec v_3 = \vec g_4 + \vec g_3^0,
\eeq 
where $\vec g_3^0 = m_3 n_3 \vec v_3$ is the momentum density carried by $^3$He atoms, not the fully dressed quasiparticles.      In the absence of $^4$He and $^3$He mass flow, $\vec v_s = -(\rho_{ph}/\rho_s)\vec v_{ph}$;  at the temperatures and concentrations of interest, $v_s \ll  v_{ph}$.

 The linearized superfluid acceleration equation is
\beq
m_4\frac{\partial \vec v_s}{\partial t} +\vec\nabla \mu_4 = 0,
\label{accel}
\eeq
where $\mu_4$ is the $^4$He chemical potential;
note that in the absence of superfluid acceleration,  $\mu_4$ is independent of position. 
In addition the hydrodynamic equation for conservation of total momentum in the dilute solutions, is, in linear order,
\beq
\frac{\partial \vec g}{\partial t} +\vec\nabla P = \vec H,
\label{navier}
\eeq
where $P$ is the total pressure and $\vec H$ is the gradient of the viscous stress tensor.  In the phonon regime, the
phonon first viscosity  is dominant; indeed the  phonon contributions to the dissipative second viscosity terms in Eqs.~(\ref{accel}) and (\ref{navier}) vanish \cite{khalat1}.   Similarly, the viscosity of the $^3$He gas can be neglected, as can small second viscosity effects associated with heat transfer between the $^3$He and phonons.  
Then
\beq
\vec H \equiv \eta_{ph} \left(\nabla^2 \vec v_{ph} +\frac13 \vec\nabla(\vec \nabla\cdot\vec v_{ph})\right),
\eeq
where $\eta_{ph}$ is the first viscosity of the phonon fluid.

The $^3$He momentum density obeys \cite{BS68}
\beq
\frac{\partial}{\partial t}\left( \vec g_3 - \delta m n_3 \vec v_s \right) + n_3\vec\nabla \mu_3 +S_3\vec\nabla T
  = \frac{\partial \vec g_3}{\partial t}\Big|_{coll},
\label{g3cons}
\eeq
where $\vec g_3$ is the momentum density of the dressed $^3$He quasiparticles,  
\beq
  S_3 = \left(\frac52 -\beta(\mu_3-\epsilon_0)\right)n_3= \left[\frac52 - \ln\left(\frac{n_3\lambda_{th}^3}{\nu}\right)\right]n_3.
\eeq
 is the $^3$He entropy density, and 
$(\partial\vec g_3/\partial t)|_{coll}$ is the rate of change of the $^3$He momentum density resulting from collisions with phonons.  This equation can also be written in terms of the $^3$He partial pressure, $P_3=n_3T$, if one uses the relation
$dP_3 = S_3 dT + n_3d(\mu_3 - \epsilon_0)$.

Similarly the phonon momentum density obeys
\beq
\frac{\partial}{\partial t}(\rho_{ph}(\vec v_{ph} - \vec v_s)) + S_{ph}\vec\nabla T   =  - \frac{\partial \vec g_3}{\partial t}\Big|_{coll}
+ \vec H,
\label{phmomcons}
\eeq 
where
\beq
  S_{ph} = s^2\rho_{ph}/T =  =\frac{2\pi^2}{45}\left(\frac{T}{s}\right)^3
\eeq
is the $^4$He  entropy density;
this equation can similarly be written in terms of the phonon partial pressure,  $P_{ph}= TS_{ph}/4$ if one uses the relation $dP_{ph} = S_{ph}dT  - \frac 34 s\rho_{ph} ds$.

With the Gibbs-Duhem relation, 
\beq
\vec\nabla P = n_4 \vec\nabla \mu_4 + n_3 \vec\nabla \mu_3 + S \vec\nabla T,
\label{eqn:GibbsDuhem}
\eeq
conservation of momentum implies 
\beq
\rho_{ph}\frac{\partial \vec v_{ph}}{\partial t} + m^* n_3\frac{\partial \vec v_{3}}{\partial t} + n_3 \vec\nabla \mu_3 
+\frac{\rho_{ph} + \delta m \,n_3}{m_4} \vec\nabla \mu_4
+ S \vec\nabla T =  \vec H;
\label{momcons2}
\label{eqn:overallTime}
\eeq
here $S$ is the total entropy density. equal at low temperatures to $S_{ph}+S_3$.
In a steady state, 
\beq
  n_3\vec\nabla \mu_3 + S \vec\nabla T  = \vec H;
  \label{dndt}
\eeq
when the phonon viscosity contribution is negligible, a condition realized at $^3$He relative concentrations $ \ga\, 10^{-4}$, a temperature gradient inevitably accompanied by a $^3$He chemical potential gradient and vice versa. 
\subsection{Energy conservation}

     Using Eqs.~(\ref{3cons})-(\ref{momcons2}), together with the relation for the internal energy $dE_{int} = TdS + \mu_4 dn_4 + \mu_3 dn_3$, we readily find the equation for conservation of energy,
\beq
  \frac{\partial E}{\partial t}  + \vec\nabla \cdot\vec j_E = 0,.
\eeq
where the energy current is 
\beq
\vec j_E = \vec g_4\mu_4/m_4 + n_3\vec v_3\mu_3+ T\vec v_3 S_3 +T \vec v_{ph} S_{ph}  +\vec Q_3. 
\label{je}
\eeq
(For the moment we omit the usual first viscosity terms in $\vec j_E$.)  For a purely linear phonon dispersion
relation the total energy transported by the phonons is simply the drift term $TS_{ph}\vec v_{ph}$, and $\vec Q_3$ is the $^3$He
heat current. 
Similarly the equation for entropy flow is
\beq
  \frac{\partial S}{\partial t}  + \vec\nabla \cdot\vec j_S = -\frac{1}{T}\vec u\cdot \frac{\partial \vec g_3}{\partial t}\Big|_{coll}
  - \vec Q \cdot \frac{\vec\nabla T}{T^2}  \equiv {\cal R},
  \label{entropycons}
\eeq
where
\beq
  \vec u \equiv \vec v_3 - \vec v_{ph};
\eeq
here the entropy current is $\vec j_S = \vec v_{ph} S_{ph} + \vec v_3 S_3 +\vec Q_3 /T $.   The quantity ${\cal R}$ is the rate of entropy generation, which must be non-negative.

   In general the change of $^3$He momentum density is driven by gradients in the chemical potential difference $\mu$ and the temperature \cite{LLFM59}, as   
we see by subtracting $y \equiv m^*n_3/(\rho_{ph} + m^*n_3)$ times Eq.~(\ref{phmomcons}) from $(1-y)$ times Eq.~(\ref{g3cons}):
\beq
   (1-y)n_3\left[m^*\frac{\partial \vec u}{\partial t}  + m_3\vec \nabla \mu\right] - \sigma \vec\nabla T = 
    \frac{\partial \vec g_3}{\partial t}\Big|_{coll} -y\vec H.
    \label{dudt}
\eeq
Here
\beq
  \mu = \frac{\mu_3}{m_3} - \frac{\mu_4}{m_4}. 
\eeq
and
\beq
\sigma = yS_{ph} - (1-y)S_3.
\eeq

    Relative motion of the $^3$He and phonons can in general generate a $^3$He heat current, a type of ``thermoelectric"
effect; thus as a function of $\vec\nabla T$ and $\vec u$, the $^3$He heat current to linear order has the form:
\beq
    \vec Q_3  = -K_3 \vec\nabla T + T\chi \vec u.
 \label{Tu}   
 \eeq
In addition, the rate of momentum transfer in $^3$He-phonon collisions has the form:
 \beq
 \frac{\partial \vec g_3}{\partial t}\Big|_{coll} =  -\frac{m^*n_3}{\tau}  \vec u- \chi \vec \nabla T, \label{uT}
 \label{g3chi}
 \eeq 
where the lifetime $\tau$ determines the diffusion constant, 
\beq
D= T\tau/m^*
\eeq
 for $^3$He against phonons.    
 That the same off-diagonal thermoelectric coefficient $\chi$ appears in both $\vec Q$ and $(\partial \vec g_3/\partial T)_{coll}$ follows from general Onsager reciprocity relations for transport coefficients \cite{LLFM596}.   As we shall see, $\chi$ vanishes for dilute solutions at very low temperature; the first non-vanishing result for $\chi$ appears when we take into account phonon recoil corrections (Sec. \ref{sec:recoil}).
    
With Eqs.~(\ref{Tu}) and (\ref{uT}), together with (\ref{dudt}) in the static limit, we find the rate of entropy generation, 
\beq
   {\cal R} = \frac{n_3}{D}u^2 + K_3\left(\frac{\vec \nabla T}{T}\right)^2,
\eeq
plus the usual first viscosity term.
 The kinetic coefficients $D, K_3$, $\chi$ and $\eta_{ph}$, specify the transport properties of the solutions;
the task we pursue in the following sections is to calculate these coefficients in terms of microscopic scattering processes.

 Since in a steady state, $n_3 \vec\nabla \mu_3 + S_3\vec\nabla T = (\partial g_3/\partial T)_{coll}$, we find from Eq.~(\ref{uT}),
 \beq
   \vec u =- \frac{D}{T}\left(\vec\nabla \mu_3 + \frac{S_3+\chi}{n_3}\vec\nabla T\right).
   \label{umut}
 \eeq  
It will be more useful in later calculations of the transport properties to work in terms of the $^3$He density, $n_3$,  rather than $\mu_3$.  Using $n_3d\mu_3 = d(n_3T) - S_3dT + n_3d\epsilon_0$,
we equivalently have
\beq
 \vec u = -D\left(\frac{\vec \nabla n_3}{n_3} + \frac{\vec \nabla \epsilon_0}{T}\right)  -  D_T\frac{\vec \nabla T}{n_3},
  \label{uLLa}
\eeq
where $D_T$, an effective thermoelectric diffusion constant, is given by 
\beq
     D_T = \frac{D}{T}(n_3+\chi).
      \label{Dchi}
\eeq

In the situation in which there is no net $^3$He or $^4$He particle flow, the total energy current, from Eq.~(\ref{je}),  is 
\beq
  \vec Q =  \vec Q_3  - TS_{ph}\vec u  = -K_3 \vec \nabla T - T(S_{ph}-\chi)\vec u;
  \label{Qtot}
\eeq
the $TS_{ph}\vec u$ term is the heat carried by phonons with respect to the $^3$He.   
We define the total thermal conductivity, $K$, of the solutions by calculating, in the absence of $^3$He and $^4$He particle transport,
the total energy flow,  $\vec Q = \vec Q_3 - TS_{ph} \vec u \equiv -K \vec \nabla T$.  In the absence of phonon viscous effects (which we include below), the Gibbs-Duhem relation implies in this situation that $n_3\vec\nabla\mu_3 =- S\vec\nabla T$, so that
in the static limit Eq.~(\ref{umut}) yields 
\beq
    \vec u = \frac{D}{n_3T}(S_{ph} - \chi)\vec\nabla T,
    \label{u1}
    \eeq
and thus
\beq
  K= K_3 + \frac{D}{n_3}\left(S_{ph} - \chi\right)^2.
  \label{Ktot}
\eeq
At constant pressure, the total entropy generation rate, ${\cal R}$, is then $K\left(\vec \nabla T/T\right)^2$.

\subsection{Phonon viscosity}

   We now extend the previous discussion of the thermal conductivity to include the phonon viscosity.
 We consider a steady state in which $\vec v_3 =0$, and assume that $\vec v_{ph}$ is in the z-direction, but
varying sinusoidally in the x-direction,  i.e., $\vec v_{ph} \sim e^{ik_x x}\hat z v_{ph}$.
Equation~(\ref{phmomcons}) for conservation of momentum thus reads
\beq
  S_{ph}\vec\nabla T = - \frac{\partial \vec g_3}{\partial t}\Big|_{coll} -\eta_{ph}k_x^2 \vec v_{ph}. 
 \label{momtot1}
\eeq
Using Eq.~(\ref{g3chi}), we find
\beq
  \vec v_{ph} = -\frac{S_{ph} -\chi}{ (Tn_3/D) + \eta_{ph} k_x^2}\vec\nabla T.
  \label{vphvisc}
\eeq
Including the term $-T\chi\vec v_{ph}$ in the total heat current $\vec Q$, we then derive the total thermal conductivity of the solution,
\beq
   K = K_3 + T\frac{(S_{ph} -\chi)^2}{ (Tn_3/D) + \eta_{ph}k_x^2} \equiv K_3 + K_{ph}. 
  \label{Keta}
\eeq
This equation defines the phonon thermal conductivity, $K_{ph}$.

   Equation~(\ref{Keta}), which encompasses the range of $x_3$ from that in the Lamoreux experiment to that in the nEDM experiment, yields a number of physically interesting limits.   Since $\eta_{ph} = \frac15 s\rho_{ph}\ell$, where $\ell$ is the phonon mean free path for viscosity (see Eq.~(\ref{etaph})), the denominator in Eq.~(\ref{Keta}) shows the competition between the mean free path for phonon scattering against the $^3$He and the viscous diffusion length $\sim 1/k_x^2 \ell$.
First, for non-zero $n_3$, as $k_x \to 0$, corresponding to a container infinite in all directions, we derive Eq.~(\ref{Ktot}).
On the other hand, in the absence of  $^3$He,
\beq
 K \to K_{ph} = \frac{TS_{ph}^2}{\eta_{ph} k_x^2} =  5\frac{sS_{ph}}{k_x^2\ell}.
  \label{53}
\eeq
This result can be understood in terms of a phonon undergoing a random walk of $\sim (1/k_x \ell)^2$ steps in going a wavelength $\sim 1/k_x$.  

    For Poiseuille flow along the z-direction between parallel plates separated at $x= \pm L_x/2$, one has $v_{ph} \sim L_x^2 - 4x^2$, so that in terms of $\bar v_{ph}$, the average flow velocity, 
$\nabla^2 v_{ph}  = - 12 \bar v_{ph}/L_x^2$.     Thus in linear order 
$
 K_{ph} =   5sS_{ph} L_x^2/12\ell,
$ 
which agrees with Eq.~(\ref{53}) with the identification $k_x^2 = 12/L_x^2$.  For Poiseuille flow in a cylinder of radius $R$  one has rather $\nabla^2 v_{ph}  = -8\bar v_{ph}/R^2$ on average, which leads to the
Benin-Maris result \cite{BM}
$
K_{ph} =   5sS_{ph} R^2/8\ell$.

   To see the regime in which phonon viscosity is important we note that the ratio of the viscous to diffusive term in the
denominator of Eq.~(\ref{vphvisc}) is 
\beq
    \frac{\eta_{ph}k_x^2D}{Tn_3} = \frac{\ell k_x^2D}{5s}\frac{S_{ph}}{n_3}.
 \eeq
 As extracted from viscosity measurements~\cite{maris,greywall}, the characteristic phonon mean free path for viscosity in pure $^4$He is
\beq
\ell (T) \simeq 3.2\times 10^{-3}/T_K^5 \,{\rm cm}
\label{greywallvisc}
\eeq 
over a broad range of temperatures around $T = 0.5$~K; here $T_K$ is the temperature measured in Kelvin.   With Eq.~(\ref{diff}) for $D$ we find 
\beq
    \frac{\eta_{ph}k_x^2D}{Tn_3} \sim 0.5 \left(\frac{10^{-6}}{x_3}\right)\left(\frac{0.45K}{T}\right)^9\left(\frac{1\, \rm cm}{R}\right)^2,
 \eeq
indicating that viscosity becomes important for $x_3\, \la\, 10^{-6}$ at temperatures of order 0.5 K.  At these low concentrations,
effects of the $^3$He on the phonon viscosity are negligible.

  As discussed in Sec.~\ref{sec:results}, $K_3$ reaches a maximum fraction of only about 1\% of the overall thermal conductivity at the highest $^3$He concentrations considered here, $x_3 = 10^{-3}$.  Although the phonon thermal conductivity $K_{ph}$ falls with increasing $x_3$ as $1/x_3$ in this regime, $K_3$ is limited by $^3$He-$^3$He scattering and is never dominant. 

\subsection{Currents and distribution functions}

At low temperatures,  the $^4$He energy current is given in terms of the  phonon distribution function, $n_{\vec q}$, by
\beq
 \vec j_{E,4} =  \left(n_4 -\frac{\delta m}{m_4} n_3\right)\mu_4 \vec v_s +  \int\frac{d^3q}{(2\pi)^3}sq (s\hat {\vec q})n_{\vec q}, 
 \eeq
where the second term is the phonon energy current, 
\beq
 \vec j_{E,ph} =  s^2\int\frac{d^3q}{(2\pi)^3} \vec q \, n_{\vec q} = TS_{ph}\vec v_{ph}.
 \label{qph}
\eeq 
The integral is simply the momentum density carried by the phonons.  The corresponding $^3$He energy current is
similarly given in terms of the $^3$He distribution function,  $f_{\vec p}$, by
\beq 
  j_{E,3}= \frac{\delta m\,n_3}{m_4} \mu_4\vec v_3 +
 \nu\int\frac{d^3p}{(2\pi)^3}\frac{\vec p}{m^*}\left(\frac{p^2}{2m^*}+\epsilon_0\right)f_{\vec  p},
 \eeq
with the $^3$He flow velocity, $\vec v_3$, defined by
\beq
 \nu\int\frac{d^3p}{(2\pi)^3}\frac{\vec p}{m^*}f_{\vec p}  =  n_3\vec v_3.
\label{v3}
\eeq
The  $^3$He heat current is 
\beq
\vec Q_3  = \nu \int\frac{d^3p}{(2\pi)^3}\frac{\vec p}{m^*}\left(\frac{p^2}{2m^*}+\epsilon_0\right)f_{\vec  p}\, -(\mu_3n_3+TS_3)\vec v_3 \nonumber\\
= \nu \int\frac{d^3p}{(2\pi)^3}\frac{\vec p}{m^*}\left(\frac{p^2}{2m^*} -\frac52 T\right)f_{\vec  p}, 
\label{q3a}  
\eeq  
since for classical statistics, $(\mu_3-\epsilon_0)n_3 + TS_3 = (5/2) n_3T$.

\section{Microscopic Scattering}
\label{sec:scattering}

The transport properties of the dilute solutions are determined microscopically by four scattering processes: $^3$He-phonon scattering, similar to the scattering of non-relativistic electrons and photons; phonon-phonon scattering, which tends to bring the phonons into equilibrium (although incompletely in the present situation), $^3$He-$^3$He scattering, 
and lastly, scattering of phonons from the walls (which we consider in Ref.~\cite{BBPlowx}).

\subsection{$^3$H\lowercase{e}-$^3$H\lowercase{e} scattering}

The matrix element for  $^3$He-$^3$He scattering for atoms of opposite spin for small momentum transfers is \cite{BE67}
\beq
V_0= -0.064 \frac{m_4s^2}{n_4}.
\eeq
Thus the $^3$He-$^3$He scattering length is
\beq
a=\frac{m^*V_0}{4\pi \hbar^2},
\eeq
the differential cross section is 
$d\sigma/d\Omega =a^2$,
and the total cross section is 
\beq
\sigma_{33 }&=&4\pi a^2 
 = \frac{m^{*2}}{4\pi\hbar^4}|V_0|^2  \nonumber \\
 &&= \frac{9\pi^3}{k_D^2}(0.064)^2\left(\frac{m^*}{m_4}\right)^2\left(\frac{m_4s}{\hbar k_D}\right)^4 
\simeq 10.5 \rm\AA^2.  \nonumber \\
\eeq 
 A $^3$He atom of low momentum scatters only from atoms of opposite spin, so that the mean free path of a $^3$He through a gas of unpolarized $^3$He is given by
\beq
\ell_{33}=\frac{2}{n_3 \sigma_{\rm tot}} = \frac{8\pi n_4}{(0.064)^2x_3}\left(  \frac{m_4}{m^*}\right)^2\left(\frac{\hbar}{m_4s}\right)^4
 \approx \frac{8.66\times 10^{-8}}{x_3}\, {\rm cm}.
\label{l33} 
\eeq

\subsection{$^3$H\lowercase{e}-phonon scattering}

For $^3$He thermal velocities, $\sim \sqrt{3T/m^*}$, small compared to $s$, the scattering is sufficiently elastic that one can, to a good approximation in calculating the $^3$He-phonon scattering matrix element,  neglect the energy transfer in a collision.   As shown in \cite{BE67} [Eq.~(24) there], the effective matrix element for elastic scattering of a phonon from momentum $\vec q$ to $\vec q\,'$ and a $^3$He from momentum $\vec p$ to $\vec p\,'$ is 
\beq
\langle p'q'|T|pq\rangle \equiv \langle {\cal T} \rangle = \frac{s\sqrt{qq'}}{2n_4\Omega}(A+B\cos\theta_{qq'}),
\label{t}
\eeq 
where $\theta_{qq'}$ is the angle through which the phonon is scattered,  the angle between $\vec q$ and $\vec q'$; 
the coefficients are parameters of a solitary $^3$He in $^4$He, deduced from experiment to be 
$A= n_4 d\alpha/dn_4 = -1.2\pm0.2$ \cite{BM67,WRR69},
 $B= (1+\alpha +\delta m/m_4)(m_4/m^*)(1+\alpha - m_3/m_4) = 0.70 \pm 0.035$ \cite{BM67,WRR69,A69}, and $\Omega$ is the volume of the system.  
 The largest uncertainty is in $A$, owing to a systematic difference between the measurements \cite{BM67,WRR69} of the pressure dependence of the density of dilute solutions.

 The differential rate of scattering of a phonon of momentum $q$ by angle
$\theta_{qq'}$ is thus
\beq
  n_3 \frac{d\gamma_q(\theta)}{d\cos\theta_{qq'}} 
    &=& n_3\int \frac{q'^2 dq'}{2\pi^2} 2\pi\delta(sq-sq') |\langle p'q'|T|pq\rangle|^2   
  \nonumber\\ 
  &=& \frac{x_3sq^4}{4\pi n_4}(A+B\zeta)^2,
\eeq  
where  $\zeta =  \cos\theta_{qq'}$; the 
transport scattering rate for phonons colliding on $^3$He atoms is  
 \beq  
  \gamma_q &=& \int \frac{d^3q'}{(2\pi)^3} 2\pi\delta(sq-sq') |\langle p'q'|T|pq\rangle|^2  (1-\cos\theta_{qq'}) \nonumber\\
  &&= \int_{-1}^1 \frac{d\zeta}{2} \frac{d\gamma_q(\theta)}{d\zeta}(1-\zeta)   = s \frac{q^4J}{4\pi n_4^2},
\eeq
where  $J = A^2 + (B ^2- 2AB)/3 \simeq 2.2 \pm 0.6$.
The momentum dependent mean free path of a phonon scattering against the $^3$He is
\beq
  \ell_{ph3}(q) = \frac{s}{n_3\gamma_q} =  \frac{5.52}{3\pi  x_3k_d }\left(\frac{k_d}{q} \right)^4,
  \label{k1}
\eeq  
where $n_4 = k_D^3/6\pi^2$ and $k_D\simeq  1.089$ \AA.

  Similarly, as we shall see  (Eq.~(\ref{son0}) with $n=0$), the effective $^3$He-phonon relaxation rate in diffusion is $\Gamma/3m^*T$, where
\beq
  \Gamma &=& \int\frac{d^3q}{(2\pi)^3} q^2 \gamma_q n_q^0(1+n_q^0) \nonumber\\
   &=& \frac {9!}{ 2} \pi\zeta(8)J \left(\frac{T}{\hbar sk_D}\right)^9 k_D^3s, 
   \label{Gamma}
\eeq 
with $\zeta(8) = \pi^4/9450 \simeq 1.004$ the Riemann zeta function.   The mean free path for scattering of a thermal $^3$He  by phonons \cite{nosculpa} is \beq
  \ell_{3ph} &=&   \frac{2}{\sqrt3\pi 8!\zeta(8)Jk_D}\left( \frac{ms}{k_D}\right)^{1/2} \left(\frac{\hbar s k_D}{T}\right)^{15/2}  \nonumber \\
      &=& 0.077 \left(\frac{0.45K}{T}\right)^{15/2} \, {\rm  cm}.
 \label{l3ph}     
\eeq  
    
 Comparing the
mean free paths, Eqs.~({\ref{l33}) and (\ref{l3ph}) to estimate the importance of 3-3 versus phonon scattering in bringing the $^3$He into equilibrium. we find
\beq
\frac{\ell_{3ph}}{\ell_{33}} =   0.89 \times 10^6 x_3  \left(\frac{0.45K}{T}\right)^{15/2}
\eeq 
For $T=0.45$ K and $x_3=10^{-6}$, $\ell_{3ph} \approx \ell_{33}$, while for 
$T=0.65$ K and $x=3\times 10^{-4}$,  $\ell_{3ph}/\ell_{33}
\approx  16.9.$

\subsection{Phonon-phonon scattering}

Phonon-phonon scattering conserves total phonon momentum, and thus does not contribute to the thermal conductivity directly.  However, its effect on the phonon distribution is important.   Because of the anomalous dispersion of phonons
in liquid $^4$He at low pressure, three-phonon Beliaev and Landau-damping processes are
allowed; these processes rapidly equilibrate phonons propagating in a given direction. 
producing a phonon distribution,  
\beq
  \tilde n_{\vec q} = \frac{1}{e^{\tilde\beta(\hat q, \vec r)sq} -1},
 \label{rays}
\eeq
along rays in momentum space \cite{maris}, in which the temperature is dependent on the momentum direction $\hat q$. 
From energy conservation in these rapid scatterings, 
\beq
    \int sq (q^2dq)  \left(n_{\vec q}  - \tilde n_{\vec q}\right) = 0.
    \label{rayeq}
\eeq
  
For phonons with small momentum $q \sim T/s$, the dominant three-phonon process
is Landau damping, in which the phonon is absorbed or emitted by a thermal phonon, thereby producing another phonon with energy $\sim T$.  The rate of this process is
\beq
  \frac{1}{\tau_L} = \frac{3\pi}{2}(u+1)^2 sq \frac{\rho_{n}}{m_4n_4},
\eeq 
where $u= \partial \ln s/ \partial \ln n_4 \simeq 2.843$ at SVP is the phonon Gr\"uneisen parameter.   The Beliaev process, the decay of one phonon into two, has a rate $(u+1)^2 q^5/240\pi m_4n_4$; this process dominates at large $q$  ($\gg T/s$) \cite{processes}.
The ratio of the rates of phonon scattering on $^3$He to Landau damping is
\beq
     \frac{15xJ}{(u+1)^2}   \left(\frac{sq}{T}\right)^3  \frac{m_4s^2}{T}  \simeq 0.5 x \left(\frac{sq}{T}\right)^3  \frac{m_4s^2}{T} \ll 1.
     \label{3phcfphph}
\eeq   
 We will describe such rapid scatterings in the phonon Boltzmann equation in terms of a relaxation time, $\tau_r$, with a collision rate,
\beq
\left(\frac{\partial n_{\vec q}}{\partial t}\right)_{rapid} = -\frac{1}{\tau_r}  \left(n_{\vec q}  - \tilde n_{\vec q}  \right).
\eeq

 Relaxation between rays, which conserves phonon momentum and energy, occurs on a longer time scale.  
We describe such processes, which determine the phonon viscosity, in a relaxation time approximation
\beq
\left(\frac{\partial n_{\vec q}}{\partial t}\right)_{long} = -\frac{1}{\tau_l}  \left(n_{\vec q}  -  n^{le}_{\vec q}  \right).
\eeq
where $\tau_l= \ell (T)/s$.   Here
\beq
  n_{\vec q}^{le}  = \frac{1}{e^{\beta(\vec r)  ( sq- \vec q \cdot \vec v_{ph})} -1}
  \label{nqle}
\eeq
is the local equilibrium phonon distribution function, in terms of the local temperature $T(\vec r) = 1/\beta(\vec r)$ and mean flow velocity $\vec  v_{ph}$ that the phonons would have by relaxing to local equilibrium through phonon-phonon collisions. 

 When phonon viscosity is important, the angular dependent temperature of the phonon distribution has the form
 \beq
  \delta \beta (\hat q)  = -\frac{\beta}{s} \left(\hat q_z v_{ph} +  \hat q_z \hat q_x \lambda \right), 
  \label{y}
\eeq 
where $\lambda$ measures the quadrupolar distortion of the phonon distribution, proportional to the 
gradient in the transverse direction of the phonon velocity $v_{ph}$, which we take along the x direction to be specific.  The relaxation time, $\tau_\ell$ depends strongly on the angular dependence of $n_{\vec q}$,  which is a second spherical harmonic (the  final term in Eq.~{{\ref{y}})) for phonon viscosity.  For $\ell \gg 1$ one expects $1/\tau_\ell \sim \ell^2$, until $1/\ell \sim \theta$, where $\theta$ is a typical scattering angle.

\section{ $^3$He Boltzmann Equation}
\label{sec:3HeB}

  We turn now to calculating the transport coefficients, $D$, $K_3$, and $\chi$ from the $^3$He Boltzmann equation.    We assume quite generally that the driving terms, $\vec \nabla \mu_3$ and $\vec \nabla T$  and thus $\vec v_{ph}$ are in the z direction.  The $^3$He Boltzmann equation has the form
 \beq
\frac{\partial f_{\vec p}}{\partial t}  +\frac{\vec p}{m^*}\cdot \vec\nabla_r f_{\vec p} 
&&=  -\frac{f_{\vec p} - f_{\vec p}^{le} }{\tau_{33}} 
\nonumber\\&&
 +\sum_{p',q,q'}|\langle{\cal T}\rangle|^2 2\pi\delta(\Delta E)\delta_{\vec p +\vec q, \vec p\,' +\vec q\,'} \nonumber\\&& \hspace{-2cm}\times\left[f_{\vec p\,'}n_{\vec q\,'}(\vec r\,)(1+n_{\vec q}(\vec r\,))-f_{\vec p}\,n_{\vec q}(\vec r\,)(1+n_{\vec q\,'}(\vec r\,)\right], 
 \label{3BE}
\eeq
where $\tau_{33}$ is the $^3$He-$^3$He scattering time (appropriate for vector drivers of the $^3$He away from equilibrium);  we take a mean thermal velocity to define $\tau_{33} = \sqrt{m^*/3T} \ell_{33}$ in terms of the $^3$He-$^3$He mean free path.
Also
\beq
f_{\vec p}^{le}  = e^{-\beta(\vec r)(p^2/2m^*  -\vec p\cdot \vec v_3- \mu_3(\vec r))} 
\eeq
is the distribution function towards which $^3$He-$^3$He collisions drive the $^3$He, and  $\langle{\cal T}\rangle \equiv \langle p'q'|T|pq\rangle$.     In calculating  $D$, $K_3$, and $\chi$, phonon viscosity can be neglected.
Then, as argued above, rapid phonon equilibration along rays in momentum space brings the phonon distribution into the form (\ref{nqle}).  We first linearize the phonon-$^3$He-phonon collision term in terms of deviations from the phonon and $^3$He distribution functions from equilibrium:
 \beq
    \delta \tilde n_{\vec q} = \tilde n_{\vec q} - n_q^{le0} =
   \beta n_q^0(1+n_q^0)sq \vec q\cdot \vec v_{ph},
\eeq
 and 
\beq
  \delta f_{\vec p} = f_{\vec p} - f_p^{le0}  = \beta f_p^0 p_z w_p;
\eeq
here the local equilibrium $^3$He distribution function is
\beq
f_p^{le0} = e^{-\beta(\vec r)(p^2/2m^* - \bar\mu_3(\vec r))}
  \label{le}
\eeq
and the global equilibrium distribution function is
\beq
 f_p^{0} = e^{-\beta(p^2/2m^* - \bar\mu_3)},
\eeq
where $\bar \mu_3 = \mu_3 - \epsilon_0$.
With the detailed balance condition,  $f_{p'}^0n_{q'}^0(1+n_q^0) = f_p^0n_q^0(1+n_{q'}^0)$,  the term in square brackets in Eq.~(\ref{3BE}) becomes in linear order,
\beq
- f_p^0n_q^0(1+n_{q'}^0)\left[ \beta(w_pp_z-w_{p'}p_z'  +(\vec q - \vec q')\cdot \vec v_{ph} \right].
 \label{lindistr}
\eeq 
  
  It is most convenient to use the momentum conservation to eliminate $\vec p\,'$, and write $\vec p\,' = \vec p + \vec k$,
where $\vec k = \vec q - \vec q\,'$. Then energy conservation implies that 
\beq
\Delta E \equiv  sq' -sq + \frac{\vec p\cdot\vec k}{2m^*} + \frac{k^2}{2m^*} = 0.
\eeq 
The momentum transfer $k$ is characteristically of order $T/s$, which is 
small compared with the momentum of a $^3$He atom, $\sim (m^*T)^{1/2}$.
The approach we take will be to expand the collision integrals in powers 
of $k/(m^*T)^{1/2}$; the leading terms are of second order.   One 
might have imagined that to leading order one could neglect the energy 
transfer in collisions.  However, this is not the case because, in a 
collision, the relative changes in the momentum and energy of a $^3$He 
atom are both of order $(T/m^*s^2)^{1/2}$.
  As we show, the differential equation we derive for 
the distribution function to order $k^2/m^*T$, in the form of a Fokker-Planck 
equation, has an exact analytic solution.

Our calculations show that the energy transfer in a collision of $^3$He with a phonon, although relatively small, has a large qualitative effect on the $^3$He distribution function.  In Ref.~\cite{highx} we showed that, if the  energy transfer were neglected, the relaxation time for a quasiparticle varied as $p^2$, whereas the exact solution above shows that when the energy transfer is taken into account, the relaxation time is independent of $p$.

The next higher-order terms vary as $k^4$ and we shall refer to them as 
``recoil corrections''. These give rise to contributions of order 
$T/m^*s^2$ times the leading term.  Despite the fact that $T/m^*s^2$ is 
$\sim10^{-2}$ at temperatures of order $0.5$ K, these corrections 
are not negligible because of the large numerical coefficients, as we 
demonstrated  in the Appendix of Ref.~\cite{highx}. We return to the recoil
corrections in Sec.\ \ref{sec:recoil}.

  Expanding $w_pp_z - w_{p'}p_z'$ to order $k^2$ we find
\beq  
w_pp_z& -& w_{p'}p'_z =- w_pk_z -(p_z+k_z)\vec k \cdot\vec \nabla_p w_p  -\frac12 p_z(\vec k \cdot\vec \nabla_p)^2 w_p \nonumber\\
 &=&- w_pk_z -(p_z+k_z)\vec k \cdot\hat {\vec p} \,w_p'  -\frac12 p_z\left[(\vec k \cdot\hat {\vec p})^2 (w_p''- w_p'/p) + k^2w_p'/p\right],\nonumber\\
 \label{expansion}
\eeq
where primes on the $w$'s denote derivatives with respect to $p$.

  The $^3$He-phonon collision term in Eq.~(\ref{3BE}), with the expansion  (\ref{expansion}), becomes  
\beq
\left(\frac{\partial f_p}{\partial t}\right)_{3-ph \,coll} = \beta\sum_{\vec q,\vec q'}|\langle{\cal T}\rangle|^2 2\pi\delta(\Delta E)\delta_{\vec p +\vec q, \vec p\,' +\vec q\,'}f^0_p\,n^0_q(1+n^0_{q\,'}) \nonumber \\
\times\left((w_p-v_{ph})k_z +(p_z+k_z)\vec k \cdot\hat {\vec p} \,w_p'  +\frac12 p_z\left[(\vec k \cdot\hat {\vec p})^2 (w_p''- w_p'/p) + k^2w_p'/p\right]\right).\nonumber\\ 
\eeq
The terms in the final parentheses are both linear and quadratic in $\vec k$.
We expand the energy delta function to first order in $\vec p\cdot\vec k/m^*$, and average over the direction of $\vec k$, keeping the direction of $\vec p$ fixed.
The terms in the final parentheses of order $k^2$ give 
\beq
     \frac{k^2p_z}{6}\left(4w_p'/p + w_p''\right) n_q^0(1+n_q^0)\delta(sq'-sq).
 \eeq
We symmetrize the terms of order $\vec k$ under the transformation $\vec q \leftrightarrow \vec q'$; the
argument of the delta function remains fixed, while $\vec k \to - \vec k$ and $ n_q^0(1+n_{q'}^0) \to n_{q'}^0(1+n_q^0)$.
Thus the terms of order $\vec k$ lead to 
\beq
    \frac{k^2p_z}{6m^*}\left(w_p-v_{ph} + pw_p'\right)\left(n_q^0-n_{q'}^0\right) \delta'(sq'-sq).
\eeq
When we integrate by parts in $q'$ this term becomes  
\beq
    -\beta\frac{k^2p_z}{6m^*}\left(w_p-v_{ph} + pw_p'\right)n_q^0(1+n_q^0)\delta(sq'-sq).
\eeq
Altogether then, in terms of $\Gamma$ defined in Eq.~(\ref{Gamma}), we have 
\beq
\left(\frac{\partial f_p}{\partial t}\right)_{3ph \,coll} = \frac{\beta\Gamma}{3}p_zf^0_p\left(w_p'' + 4\frac{w'_p}{p} -\frac{\beta}{m^*}\left(pw_p' +w_p-v_{ph} \right)\right).
\label{son0}
\eeq

  Remarkably, the derivatives of the distribution function in the parentheses on the right are just those entering into the equation for
the polynomials, $g_n(y)$ [with $g_0 \sim 1$, $g_1 \sim y^2-5$, etc.], which obey 
\beq
   g_n'' +(4/y - y)g'_n+ 2ng_n = 0.
\eeq
These polynomials are related to the Sonine polynomials familiar from the theory of dilute gases by the relation $g_n(y)\propto S_{3/2}^n(y^2/2)$ \cite{LP,sonine}.
Writing $w_n(p) = g_n(p\sqrt{\beta/m^*})$  we see that
\beq
   w_n'' + 4w_n'/p - (\beta/m)pw_n' = -2n(\beta/m^*) w_n.
\eeq 
so that the eigenfunctions of the differential operator in Eq.~(\ref{son0}) are just the polynomials $g_n(y)$.
The collision term for a deviation proportional to  $g_n(y)$  is thus
\beq
\left(\frac{\partial f_p}{\partial t}\right)_{3ph \,coll}& = &-\frac{\beta^2\Gamma}{3m^*}p_zf^0_p\left[(2n+1)w_n - v_{ph}\right]
\nonumber\\
&=& -\frac{\beta\Gamma}{3m^*}[(2n+1)\delta f_n - \beta f_p^0 p_zv_{ph}].
\label{son0}
\eeq

   We first consider steady state diffusion driven by a $^3$He chemical potential gradient at constant temperature.  The driving term on the left side of the Boltzmann equation is $(p_z/m^*)\beta  f_p^0\partial \bar\mu_3/\partial z$, and the solution is the first Sonine polynomial, a constant.   In other words, the $^3$He distribution is just an equilibrium drifting at velocity $v_3$, for which the $^3$He-$^3$He collision term vanishes.
Then 
\beq
 \frac{p_z}{m^*}\beta  f_p^0\frac{\partial\bar \mu_3}{\partial z} = -\frac{\beta^2\Gamma}{3m^*}p_zf^0_p u,
 \label{drift}
\eeq
where $u = v_3 - v_{ph}$, so that $u = -(3T/\Gamma) \partial \bar\mu_3/\partial z$;
the diffusion constant is thus
\beq
  D = \frac{3T^2}{\Gamma}, \label{eqn:D}
 \eeq 
and the collision time in Eq.~(\ref{g3chi}) is  $\tau = 3m^*T/ \Gamma$.
Using Eq.~(\ref{Gamma}), with  $\hbar sk_D = 19.9$K and $\hbar k_D/m_4s= 0.729$ we find numerically that
\beq
  D  = 173\left(\frac{0.45K}{T_K}\right)^7 {\rm cm^2/sec} = \frac{0.65}{T_K^7} {\rm cm^2/sec}.
     \label{diff}
\eeq
We note that there is a $\sim 35$\% increase in this lowest order calculation of $D$ due to the recoil corrections (see Secs.~\ref{sec:recoil} and \ref{sec:results}).

  We next compute the thermal conductivity of the $^3$He, defined by the $^3$He heat current produced by a temperature gradient at constant $^3$He pressure, cf. Eq.~(\ref{q3a}).   At constant $P_3$, the driving term on
the left side of the Boltzmann equation is proportional to
\beq
\left( \frac{\partial f_p^{le}}{\partial z}\right)_{P_3} = - f_p^0\left(\frac{p^2}{2m^*}  - \frac52 T\right)
 \frac{\partial \beta}{\partial z}.
\eeq
Thus the deviation of the $^3$He distribution is a first ($n=1$) Sonine polynomial $\sim p^2 - 5m^*T$;
we write
\beq
   w_p = C\left(\frac{p^2}{2m^*}  - \frac52 T\right).
\eeq
Wiith Eq.~(\ref{son0})  the Boltzmann equation yields
\beq
   C= - \frac{1}{\Gamma'}  \frac{\partial T}{\partial z},
\eeq
where $\Gamma' = \Gamma + m^*/\beta\tau_{33}$.
Thus the $^3$He heat current is
\beq
   Q_3 =  -\nu\int \frac{d^3p}{(2\pi)^3} \frac{p_z^2}{m^*}\left(\frac{p^2}{2m}  - \frac52 T\right) ^2\frac{\beta}{\Gamma'}  \frac{\partial T}{\partial z}.
 \eeq
Evaluating the integral we find the $^3$He thermal conductivity,
\beq
   K_3 = \frac{\beta}{\Gamma'}\nu\int \frac{d^3p}{(2\pi)^3}   \frac{p^2}{2m} \left(\frac{p^2}{2m}  - \frac52 T\right) ^2 = \frac{5T^2}{2\Gamma'}n_3, \label{eqn:K3}
\eeq
Numerically, the $^3$He-$^3$He scattering contribution to $\Gamma'$ is of relative order $(10^{-6}/x_3)(T/0.45K)^{15/2}$, indicating that $^3$He-$^3$He scattering does not contribute importantly
at concentrations well below $10^{-6}$ at temperatures of order 0.5 K.

   At the present level of approximation the thermoelectric coefficient $\chi$ is zero.  To see this we assume a uniform temperature; then the solution of the Boltzmann equation is simply a drifting local equilibrium [cf. Eq.~(\ref{drift})] for which, from the orthogonality of the Sonine polynomials, the $^3$He heat current vanishes.  Thus from Eq.~(\ref{Tu}), $\chi = 0$.   Similarly $(\partial g_3/\partial t)_{coll}$ also vanishes if the distribution function is proportional to the first Sonine polynomial, and thus from Eq.~(\ref{g3chi}), we see again that
$\chi$ vanishes.  However, with terms of higher order in $k$ included in the collision term, the distribution functions are not given simply in terms of Sonine polynomials, and one finds $\chi \ne 0$ (see Sec.~\ref{sec:therm}.   We expect, from expanding the $^3$He-phonon collision term to higher order in $k$ that the corrections are $\sim k^2/p^2 \sim T/m^*s^2$.  Thus in order of magnitude $\chi \sim n_3T/m^*s^2$, which leads to negligible corrections
to $D_T$, Eq.~(\ref{Dchi}), as well as $K_{ph}$ in Eq.~(\ref{Ktot}).

\section{Phonon Boltzmann Equation}
\label{sec:phononB}

  We turn to the details of the phonon Boltzmann equation.
  With phonon-phonon scatterings treated in a relaxation time approximation, the phonon Boltzmann equation in dilute solutions has the structure:
\beq
\frac{\partial n_{\vec q}}{\partial t}  +s\hat q\cdot \vec\nabla_r n_{\vec q} 
&=& \left(\frac{\partial n}{\partial t}\right)_{{\rm ph-^3He}}-\frac{1}{\tau_r}  \left(n_{\vec q}  - \tilde n_{\vec q}  \right)
- \frac{1}{\tau_l}(n_{\vec q} - n_q^{le}),\nonumber\\
 \label{phBE}
\eeq
where the phonon-$^3$He collision term is, cf. Eq,~(\ref{3BE}),
\beq
\left(\frac{\partial n}{\partial t}\right)_{{\rm ph-^3He}} = &&
\sum_{p,p',q'}|\langle{\cal T}\rangle|^2 2\pi\delta(\Delta E)\delta_{\vec p +\vec q, \vec p\,' +\vec q\,'} \nonumber\\
&&\times \left[f_{\vec p\,'}n_{\vec q\,'}(\vec r\,)(1+n_{\vec q}(\vec r\,))\right. - \left.f_{\vec p} n_{\vec q}(\vec r\,)(1+n_{\vec q\,'}(\vec r\,)\right].\nonumber\\ 
\label{3phcoll}
\eeq 
 We do not include explicit scattering of phonons with the container walls in this paper.

  In the limit that the rate of phonon collisions with the $^3$He is much smaller than phonon-phonon collisions along rays, 
cf. Eq.~(\ref{3phcfphph}), and the spatial and temporal variations are slow, the deviation of the distribution function $n_{\vec q}$ from $\tilde n_{\vec q}$ is of relative order $\tau_r$;  as we see from Eq.~(\ref{phBE}),
\beq
   n_{\vec q}-\tilde n_{\vec q} &=& \tau_r\left[ \frac{\partial n_{\vec q}}{\partial t}  +s\hat q\cdot \vec\nabla_r n_{\vec q} +\frac{1}{\tau_l}(n_{\vec q} - n_q^{le})
- \left(\frac{\partial n}{\partial t}\right)_{{\rm ph-^3He}}  \right].
\label{phbe2}
\eeq
Thus all $n$ on the right can be replaced by $\tilde n$.   The condition (\ref{rayeq}) then allows us to derive
the effective Boltzmann equation for $\beta(\hat q,\vec r)$, since 
the integral $\int dq\, q^3$ of the right side of Eq.~(\ref{phbe2}) vanishes.    We write
\beq
    \delta \tilde n_{\vec q} = \tilde n_{\vec q} - n_q^{le0} =
    -n_q^0(1+n_q^0)sq\,\delta \beta(\hat q, \vec r),
\eeq
where $\delta \beta (\hat q, \vec r) = \beta (\hat q, \vec r) -\beta(\vec r)$,
with $\beta(\vec r)$ the local equilibrium temperature,
and using
$
  \int q^3 dq \delta \tilde n_{\vec q} = -6\pi^2 s \rho_{ph} \delta \beta.
$ we arrive at the equation for $\beta(\hat q, \vec r)$:    
\beq
 \frac{\partial \beta(\hat q,\vec r)}{\partial t}  +s\hat q\cdot \vec\nabla_r \beta(\hat q,\vec r) 
&=& {\cal C} - \frac{1}{\tau_l}(\beta(\hat q,\vec r) + \beta q_z v_{ph}/s).
 \label{beC}
\eeq
Here
\beq
  {\cal C} = - \int \frac{q^3 dq}{6\pi^2 s \rho_{ph}} \left(\frac{\partial \tilde n}{\partial t}\right)_{{\rm ph-^3He}};
\eeq
the tilde on the right side indicates that the phonon distribution functions $n_{\vec q}$ in the collision terms are replaced by $\tilde n_{\vec q}$.

   The analysis of the $^3$He-phonon collision term in the phonon Boltzmann equation is similar to that in the $^3$He
Boltzmann equation.  We assume $\beta(\hat q)$ to have the form (\ref{y}), and treat the $v_{ph}$ and $\lambda$ terms separately.   In the $v_{ph}$ term, using Eq.~(\ref{expansion}), we see that when one averages over the direction of $^3$He momenta $\vec p$ only the terms first order in $\vec k$ survive;  then
\beq
\left(\frac{\partial n}{\partial t}\right)_{{\rm ph-^3He}} = &&
\sum_q  k_z |\langle{\cal T}\rangle|^2 2\pi\delta(sq-sq')
 n_q^0(1+n_q^0)) F_3 \nonumber\\
&&= q\gamma_q  n_q^0(1+n_q^0)) F_3,
\label{3phcolla}
\eeq
where
\beq
  F_3 =  \sum_p  f_p^0\left(w_p -v_{ph}+\frac13 pw_p'\right).
\eeq
If the $^3$He are driven by $^3$He chemical potential gradients, then $F_3 = n_3 u$, while if the $^3$He are
driven by a temperature gradient, then $w_p  = -(1/\Gamma)  (\partial T/\partial z) \times$ $\left(p^2/2m^* - 5T/2\right)$, and
$F_3= -n_3 v_{ph}$.

  The $\lambda$ terms in Eq.~(\ref{y}) lead to a similar term in the collision integral,
  \beq
  -\gamma_q^{(2)}\left(\delta \tilde n_{\vec q}  - n_q^0(1+n_q^0)\beta q_z v_{ph}\right),
\eeq
where 
\beq
 \gamma_q^{(2)} = \int \frac{d\cos\theta}{2}(1-P_2(\cos\theta))\frac{d\gamma_q(\theta)}{d\cos\theta} 
\eeq
is the average of the scattering rate of phonons on $^3$He appropriate for viscosity, with $P_2$ the 
second Legendre polynomial.  Thus
\beq
 {\cal C} = \frac{n_3\beta^2 \hat q_z}{3s\rho_{ph}}\left(-\Gamma u + \Gamma^{(2)}\hat q_x\lambda\right),
\eeq
where $\Gamma$ is given by Eq.~(\ref{Gamma}), and
\beq
\Gamma^{(2)} = \int \frac{d^3 q}{(2\pi)^3} q^2 \frac{\gamma^{(2)}(q)}{n_3}n_q^0(1+n_q^0) = \frac{A^2+B^2/5}{A^2+(B^2-2AB)/3}\Gamma;
\eeq
the numerical coefficient is $\simeq 0.74$.

  The linearized phonon Boltzmann equation, Eq.~(\ref{beC}), separates into equations for the individual spherical harmonic components of $\beta(\hat q)$.  We first calculate the phonon viscosity the second spherical harmonic term; to do so we assume static flow and compute the off-diagonal component $T_{xz} = -\eta_{ph} \partial v_{ph}/\partial x$.   Then
\beq
   \lambda = -s\tau_v \vec\nabla_x v_{ph}
   \label{lambda}
\eeq
where 
\beq
    \frac{1}{\tau_v} =  \frac{n_3\beta}{3\rho_{ph}}\Gamma^{(2)} +  \frac{1}{\tau_l}.
    \label{tauv}
\eeq
With Eq.~(\ref{lambda}), we find 
\beq
  T_{zx} &=&  -\int \frac{d^3q}{(2\pi)^3} sq_z\hat q_x \delta n_{\vec q} 
  \nonumber\\
  &&= \int \frac{d^3q}{(2\pi)^3}
  \beta s^2 \tau_v \hat q_z^2\hat q_x^2q^2 n_q^0(1+n_q^0)\frac{\partial v_{ph}}{\partial z},
\eeq 
and thus
\beq
  \eta_{ph} = \frac{1}{5} \rho_{ph} s^2 \tau_v.
  \label{etaph}
\eeq
The contribution to the viscosity from phonon scattering on the $^3$He is important for $(n_3/S_{ph})(s\ell/D) \simeq 3\times 10^4 x_3/T_K \,\ga \,1$. 

   In extracting the first spherical harmonic component of Eq.~(\ref{beC}), we encounter a term $ \hat q_z \hat q_x\hat q\cdot \vec\nabla_r \lambda$, which from Eq.~(\ref{lambda}) equals $ - s\tau_v  \hat q_z \hat q_x^2 \vec\nabla_x^2  v_{ph}$.
The first spherical harmonic component of $\hat q_z \hat q_x^2$ is 
$
    3 \int d\Omega(\vec q) \hat q_z^2 \hat q_x^2/4\pi = 1/5.
$
Thus the phonon momentum density obeys
\beq
  \rho_{ph}\frac{\partial  v_{ph} }{\partial t}  + S_{ph}\frac{\partial T}{\partial z} 
 -\rho_{ph}\frac{s^2\tau_v}{5} \vec\nabla_x^2 v_{ph} 
= \frac{n_3\beta} {3}\Gamma u. 
\eeq
which we recognize as the phonon momentum conservation equation (\ref{phmomcons}).

\section{Recoil corrections}
\label{sec:recoil}

As we showed above, to leading order in an expansion of the $^3$He-phonon collision operator in powers of $k^2/m^*T\sim T/m^*s^2$, the eigenfunctions of the $^3$He-phonon collision operator in the $^3$He Boltzmann equation have the form of Sonine polynomials times $p_i$ for driving terms proportional to $p_i\partial f_{\vec p}^{le}/\partial x_i$.
Terms of order $k^4/m^{*2}T^2$ in the expansion of the collision operator lead to contributions to the transport coefficients of nominal order $T/m^*s^2$ relative to the leading term.  The task in this section is to calculate these corrections explicitly.  We have seen that in a temperature or $^3$He chemical potential gradient, the phonon distribution function is one of drifting local equilibrium, and therefore, it is convenient to work in the reference frame in which the phonon fluid is at rest ($\vec v_{ph}=0$). We also assume that the system is in a steady state and, on the left hand of the Boltzmann equation (\ref{3BE}), we may replace the distribution function by a local equilibrium one.  At the low concentrations of interest, the $^3$He-$^3$He scattering may be neglected.  

   For temperature and chemical potential gradients in the z direction, the  left side of the Boltzmann equation (\ref{3BE}) becomes
   \beq
\frac{\vec p}{m^*}\cdot \vec\nabla_r f_{\vec p} =&&-\frac{ f^0_{\vec p}}{T}\frac{p_z}{m^*} \left(\frac{\partial(\mu_3-\epsilon_0)}{\partial z}+\frac{p^2/2m^*+\epsilon_0-\mu_3}{T}\frac{\partial T}{\partial z}\right)\nonumber \\
=&&-\frac{ f^0_{\vec p}}{T}\frac{p_z}{m^*} \left(\frac{1}{n_3}\frac{\partial P_3}{\partial z}-\frac{\partial \epsilon_0}{\partial z}+\frac{(p^2/2m^*-5T/2)}{T}\frac{\partial T}{\partial z}\right).
\eeq
The latter form, in which we use the $^3$He pressure and the temperature as independent variables, is particularly convenient because it shows that the driving term due to the pressure gradient is proportional to the particle current, $p_z/m^*$, carried by an atom,  and that the driver due to the temperature gradient is proportional to the heat current, $( p^2/2m^*-5T/2) p_z/m^*$.   Since the Boltzmann equation is linear, we write
$w(p)=W^c(p)\left( {1}/{n_3}{\partial P_3}/{\partial z}-{\partial \epsilon_0}/{\partial z}\right)+W^{h}(p)\partial T/\partial z$, and have the equations for the $W$'s: 
\beq
\frac{p_z}{m^*} = \,\sum_{\vec p',\vec q,\vec q'}|\langle{\cal T}\rangle|^2 2\pi\delta(\Delta E)\delta_{\vec p +\vec q, \vec p\,' +\vec q\,'} n^0_{\vec q}(1+n^0_{\vec q\,'}) 
[p_zW^c(p)-p'_zW^c(p')]. \nonumber\\
 \label{3BEsimC}
\eeq
and 
\beq
\left(\frac{p^2}{2m^*}-\frac52 T\right)\frac{p_z}{m^*}& =& \,\sum_{\vec p',\vec q,\vec q'}|\langle{\cal T}\rangle|^2 2\pi\delta(\Delta E)\delta_{\vec p +\vec q, \vec p\,' +\vec q\,'} n^0_{\vec q}(1+n^0_{\vec q\,'}) \nonumber\\
 &&\quad\quad \times[p_zW^h(p)-p'_zW^h(p')]. 
 \label{3BEsimH}
\eeq

  The response of the $^3$He particle number current density and heat current density to  the chemical potential and temperature gradients, may be written in terms of the correlation functions $\Xi_{\eta \lambda}$  between the currents as 
    \beq
  \vec j_3=-\Xi_{cc} \left(\frac{1}{n_3}\vec \nabla P_3-\vec \nabla \epsilon_0\right)-\Xi_{ch} \frac{\vec \nabla T}{T},
  \label{j3phen}
  \eeq
  and 
  \beq
  \vec Q_3=-\Xi_{hc} \left(\frac{\vec \nabla P_3}{n_3}-\vec \nabla \epsilon_0\right)-\Xi_{hh} \frac{\vec \nabla T}{T},
   \label{Q3phen}
  \eeq
where $c$ denotes the current of particle number and $h$ the heat current. 
The Onsager reciprocal relations imply that $\Xi_{ch}=\Xi_{hc}$\,, a fact that can be confirmed explicitly using the results for the $\Xi_{\eta \lambda}$ given below.

On eliminating $(\vec \nabla P_3)/{n_3}-\vec \nabla \epsilon_0$ from Eqs.\ (\ref{j3phen})  and (\ref{Q3phen}), we find
\beq
  \vec Q_3=\frac{\Xi_{hc}}{ \Xi_{cc}}n_3\vec u  -\left( \Xi_{hh}-\frac{ \Xi_{hc}^2}{ \Xi_{cc}}\right) \frac{\vec \nabla T}{T},
  \eeq
a result valid in an arbitrary frame.  Comparison of this result with Eq.\ (\ref{Tu}) shows that 
\beq
\chi=\frac{n_3}{T}\frac{\Xi_{hc}}{ \Xi_{cc}}
\label{chi_Xi}
\eeq
 and the $^3$He thermal conductivity is
$K_3= \Xi_{hh}-{ \Xi_{hc}^2}/{ \Xi_{cc}}$.

Microscopically, the number and heat current densities are given by
\beq
(j_3)_z=\nu\int \frac{d^3p}{(2\pi)^3} \frac{p_z}{m^*} \frac{ f^0_{\vec p}}{T}p_z w(p), 
\label{j3}
\eeq
and 
\beq
Q_{3z}=\nu\int \frac{d^3p}{(2\pi)^3} \left(\frac{p^2}{2m^*}-\frac52 T\right)\frac{p_z}{m^*} \frac{ f^0_{\vec p}}{T}p_z w(p), 
\label{Q3}
\eeq
and therefore
\beq
\Xi_{\eta,\lambda}= \nu\int \frac{d^3p}{(2\pi)^3} \frac{ f^0_{\vec p}}{T}X^\eta(p) \Phi^\lambda(p),
\label{Xi}
\eeq
where $X^c(p)=p_z/m^*$, $X^h(p)=(p^2/2m^* -5T/2)p_z/m^*$, and $\Phi^\lambda(p)=W^\lambda p_z$.

We may write Eqs.\ (\ref{3BEsimC}) and (\ref{3BEsimH}) in a compact matrix notation as
\beq
|X^\lambda\rangle =I|\Phi^\lambda \rangle,
\label{Boltz3symb}
\eeq
in terms of which
\beq
\Xi_{\eta,\lambda}=  \langle X^{\eta}|\Phi^\lambda\rangle,     
\label{XiCompact}
\eeq
where the inner product is defined by 
\beq
\langle C|D\rangle = \nu\int \frac{d^3p}{(2\pi)^3} \frac{ f^0_{\vec p}}{T}C^\eta(p) D^\lambda(p).
\eeq

 We now expand the collision integral in powers of $k^2/m^*T$,
\beq
I=I_0+I_1 +\ldots,
\eeq
where $I_0$ is the leading term, which is $\sim T/m^*s^2$, and $I_1$ is the term of order $(k^2/m^*T)^2\sim (T/m^*s^2)^2$.  Similarly, we write the deviation function in the form
\beq
|\Phi^\lambda\rangle =|\Phi^\lambda_0\rangle+|\Phi^\lambda_1\rangle+\ldots\,.
\eeq
Equating terms of the same order in $T/m^*s^2$ in Eq.\ (\ref{Boltz3symb}), one finds
\beq
|X^\lambda\rangle =I_0|\Phi^\lambda_0 \rangle,
\eeq
and 
\beq
0=I_0|\Phi^\lambda_1 \rangle+I_1|\Phi^\lambda_0 \rangle,
\eeq
from which one sees that
\beq
|\Phi^\lambda_1 \rangle=-(I_0)^{-1}I_1|\Phi^\lambda_0 \rangle\,.
\eeq
The transport coefficient giving the response of a variable specified by $|X^\eta\rangle$ may be expressed in the form
\beq
\Xi^{\eta,\lambda}&=&\langle X^\eta |\Phi^\lambda\rangle = \langle X^\eta|\Phi^\lambda_0\rangle +\langle X^\eta |\Phi^\lambda_1\rangle +\ldots \nonumber \\ &&= \langle X^\eta |\Phi^\lambda_0\rangle -\langle X^\eta| (I_0^{-1})I_1|\Phi^\lambda_0 \rangle +\ldots =\langle \Phi^\eta |I_0 |\Phi^\lambda_0\rangle -\langle \Phi^\eta |I_1|\Phi^\lambda_0 \rangle +\ldots  \,, \nonumber\\
\label{Xi_expand}
\eeq
from which one sees that the corrections to the transport coefficient are given by 
\beq
\Xi^{\eta,\lambda}_1= -\langle \Phi^\eta_0|I_1|\Phi^\lambda_0 \rangle\,.
\eeq

\subsection{Diffusion}
In the absence of a temperature gradient, the flux density of $^3$He atoms in the frame moving with the phonons is given by
\beq
n_3\vec u=-\Xi_{cc}\vec \nabla (\mu_3-\epsilon_0)=-\Xi \frac{T}{n_3}\vec \nabla n_3,
\eeq
and therefore $\Xi_{cc}$ is related to the diffusion coefficient by 
\beq
\Xi_{cc}=D n_3/T.
\eeq
One then sees from Eq.\ (\ref{Xi_expand}) that the fractional change in the diffusion coefficient due to recoil is
\beq
\frac{D_1(T)}{D_0(T)} \simeq -\frac{\langle \Phi^c_0|I_1|\Phi^c_0 \rangle}{\langle \Phi^c_0|I_0|\Phi^c_0 \rangle}.
\eeq
This ratio is precisely that evaluated in the Appendix of Ref.\ \cite{highx}, and therefore
\beq
\frac{D(T)}{D_0(T)} \simeq 1 +\frac{T}{m^*s^2}\left(\frac{100\pi^2-198}{33} \frac{\tilde J}{J} -1    \right)\,,
\label{D_lin}
\eeq
where ${\tilde J}=4 A^2/3 -4 AB/3+8B^2/15 \approx 3.30$ and the coefficient of $T/m^*s^2$ is $35.5$.

The reason that the calculation of recoil corrections is independent of $x_3$ is that the solution to the $^3$He Boltzmann equation is $w(p)={\rm  constant}$: for the case of low $x_3$ considered in this paper, this is the exact solution for $^3$He--phonon scattering when $^3$He-$^3$He scattering is negligible.  This is not altered at higher $x_3$ because $\Phi(p)\propto p_z$ is still a solution of the Boltzmann equation when $^3$He-$^3$He collisions are included, since they conserve the total $^3$He momentum. 

  An alternative approach to calculating recoil corrections is to start
from the standard variational expression for transport coefficients
\cite{LP}, which in the case of diffusion is
\beq
\frac{Dn_3}{T}\ge  \frac{\langle X^c |  \Phi\rangle^2}{\langle \Phi |I
|\Phi \rangle},
\eeq
for an arbitrary form of the function $\Phi$.  In particular, if one
chooses $\Phi=\Phi^c_0$ one has
\beq
\frac{Dn_3}{T}\ge  \frac{\langle X^c |  \Phi^c_0\rangle^2}{\langle
\Phi^c_0 |I |\Phi^c_0\rangle},
\label{D_var}
\eeq
which recovers the exact results for the $D_0$ and $D_1$ but also has
terms of higher order in $k^2$.

\subsection{Thermal diffusion}
\label{sec:therm}

We turn now to $\Xi^{hc}$, which describes the heat flow induced by a relative motion of phonons and $^3$He atoms (and also the relative motion of phonons and $^3$He induced by a temperature gradient).  To first order in $T/m^*s^2$ this quantity vanishes since $\langle \Phi^h_0 |I_0 |\Phi^c_0\rangle$ vanishes.  This is because $\Phi^c(p)\propto p_z$ is an eigenstate of $I_0$ and therefore $I_0 |\Phi^c_0\rangle\propto |\Phi^c_0\rangle$.  As shown in Sec.\ 4, the solution for the Boltzmann equation for response to a temperature gradient is proportional to $(p^2/2m^*-5T/2)p_z$ to leading order in $T/m^*s^2$ and consequently $\langle \Phi_0^h |I_0 |\Phi^c_0\rangle \propto \langle\Phi^h |\Phi^c\rangle$ because of the orthogonality relation for Sonine polynomials for $n=0$ and $n=1$, $\int_0^\infty dx x^4\exp(-x^2/2)(x^2-5)=0$.  The leading contributions to $\Xi^{hc}$ are therefore of second order in $T/m^*s^2$, while those to $\Xi^{cc}$ are of first order.  In addition, $\Xi^{hc}$ has an extra factor of $T$ compared with $\Xi^{cc}$ because of the extra factor $p^2/2m^*-5T/2$, and therefore as noted in Sec.~(\ref{sec:3HeB}),  the magnitude of $\chi$ is $\sim n_3T/m^*s^2$. 

In Secs.\  2.2 and 2.3 we saw that $\chi$ usually enters in combinations such as $n_3 +\chi$, $S_3+\chi$, or $S_{ph}-\chi$.  Therefore,
under the conditions of the nEDM proposed experiment, we expect the nonzero value of $\chi$ to have little effect and we shall not evaluate the prefactor.

 \section{Results}
\label {sec:results}

We now summarize the main results of this paper in graphical form.  Over the full range of $^3$He concentrations in the non-degenerate regime, from that of the Lamoreaux et al. experiment~\cite{lamoreaux} ($x_3\, \sim \,10^{-3}$) to that of the proposed nEDM experiment~\cite{snsExpt} ($x_3 \,\la \,10^{-10}$), the diffusion constant, $D$, Eq.~(\ref{diff}), is independent of $x_3$, even taking the recoil corrections of Sec.~\ref{sec:recoil} into account.  As Fig.~\ref{fig:D} shows, its basic $T^{-7}$ temperature dependence is slightly modified by the recoil corrections.  As expected, the phonon and $^3$He thermal conductivities (Eqs.~(\ref{Keta}) and (\ref{eqn:K3}), respectively) depend on both $T$ and $x_3$.  As shown in Fig.~\ref{fig:K}, at low $^3$He concentrations, the phonon thermal conductivity is dominated by phonon viscosity; at the highest concentrations considered here, $K_{ph}$ is simply proportional to 1/$x_3
$ because of phonon scattering from $^3$He.  On the other hand, the $^3$He thermal conductivity grows linearly with increasing $x_3$ at low concentrations before becoming dominated by $^3$He-$^3$He scattering at concentrations of about $x_3 \sim 10^{-6}$.  At all concentrations considered here, the $^3$He contribution to the transport of heat is $\la$ 1\% that of the phonons.  We note that we have not taken into account the effects of geometry on the mean free paths of phonons and $^3$He, especially important at low temperatures and low concentrations, respectively, in this treatment  but will do so in Ref.~\cite{BBPlowx}.

\begin{figure}
\begin{center}
\includegraphics[width=11cm]{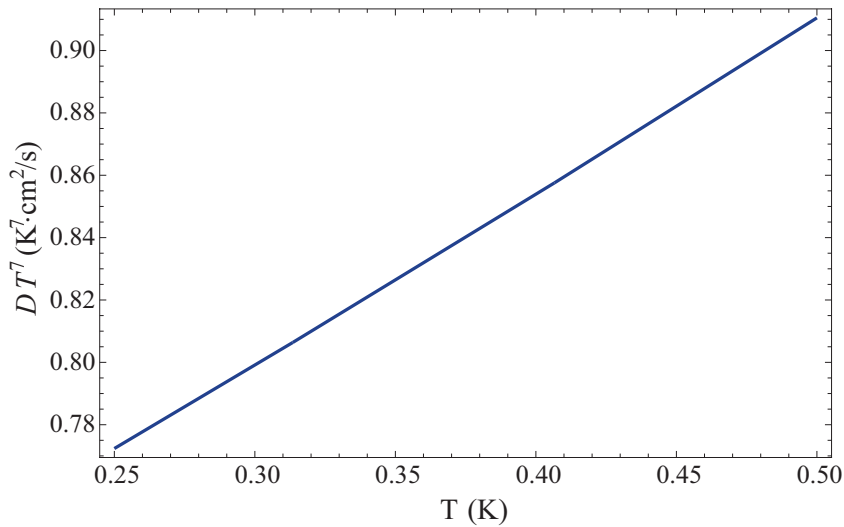}
\end{center}
\caption{ Diffusion constant multiplied by $T^7$, as a function of
temperature.  The results include recoil corrections and were obtained
from numerical integration using Eq.~(\ref{D_var}).  We note that the
expression (\ref{D_lin}), which includes only the contributions $D_0$ and $D_1$,
gives results that lie within 3\% of this result over the range shown.
\label{fig:D}}
\end{figure}

\begin{figure}
\begin{center}
\includegraphics[width=11cm]{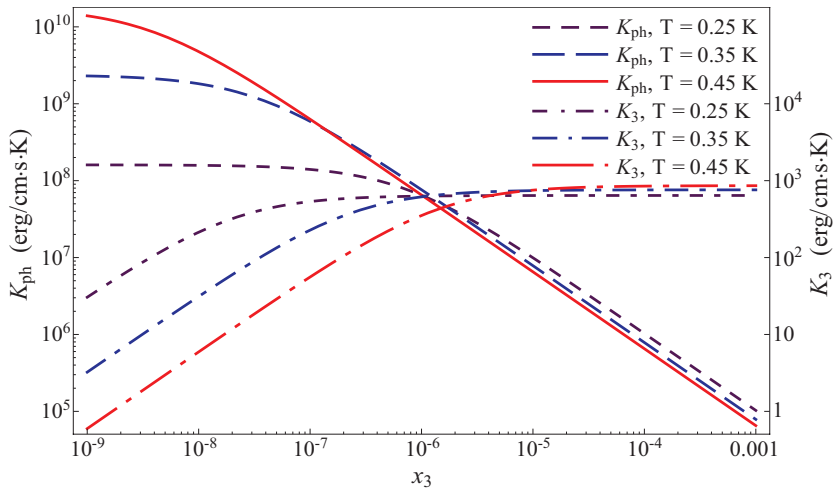}
\end{center}
\caption{Phonon and $^3$He thermal conductivities from Eqs.~(\ref{Keta}) and (\ref{eqn:K3}), respectively, at $T= 0.25$, 0.35 and 0.45 K.\label{fig:K}  Note that the scale of phonon thermal conductivities is $10^5$ larger than the scale of $^3$He thermal conductivities. The phonon thermal conductivity is calculated using the phonon mean free path, Eq.~(\ref{greywallvisc}), and $k_x^2 = 8/R^2$, with $R=15$ cm (approximately 10 times the phonon mean free path at $T=0.25$~K).  Both $K_{ph}$ and $K_3$ are calculated using the recoil correction obtained from numerical integration as described in the caption of Fig.~\ref{fig:D}}
\end{figure}

\section{Summary}
\label{sec:summary}

In this paper we have calculated the transport properties of $^3$He in superfluid $^4$He in the non-degenerate regime $x_3 \,\la\, 10^{-3}$ for
$T\, \la\, 0.6$ K where phonons are the dominant $^4$He excitations.    These calculations are relevant to previous transport measurements at relatively high concentrations \cite{lamoreaux,rosenbaum},  to the range of natural concentration, $x_3 \sim 10^{-6}$~\cite{harvard}, 
as well as to the low concentrations expected in the neutron EDM experiment~\cite{snsExpt}. We began by considering particle number, momentum and energy conservation, including the effects of phonon viscosity.   The time evolution of the relative velocity $\vec u = \vec v_3 - \vec v_{ph}$
is given by Eq.~(\ref{dudt});  the relative velocity is driven by both gradients of the chemical potentials and temperature, and includes the dissipative effects of $^3$He-phonon scattering and phonon viscosity.  We show explicitly that the entropy generation rate is positive definite.  The energy currents lead us to identify the thermal conductivities of the $^3$He ($K_3$, Eq.~(\ref{Tu})) and of the phonons ($K_{ph}$, Eq.~(\ref{Keta})).  In addition we also identify the ordinary and ``thermoelectric'' diffusion coefficients $D$ and $\chi$, Eq.~(\ref{g3chi}), as well as the Onsager reciprocity relation connecting the $^3$He heat current and the thermoelectric diffusion coefficient.

Before calculating the transport coefficients in a Boltzmann equation framework, we discuss in detail in Sec.~\ref{sec:scattering} the various scattering rates in the problem.  Whereas the $^3$He-$^3$He scattering rate is much larger than the $^3$He-phonon rate at the high concentrations of the Lamoreaux, et al.~\cite{lamoreaux} and Rosenbaum, et al.~\cite{rosenbaum}  measurements, at $x_3 \sim 10^{-6}$, they are roughly equal, and at lower concentrations, the scattering of $^3$He by phonons (or walls, depending on the geometry~\cite{BBPlowx}) dominates.  In treating phonon-phonon scattering we use the fact that phonons propagating in  a given direction in momentum are in local equilibrium, owing to the very large small angle scattering rate.   The phonon viscosity arises from scattering of phonons through large angles,
either in a single large angle event or in a sequence of small angle scatterings.   When effects of phonon viscosity are important, it is necessary to consider the effects of phonons scattering from the walls, the only appearance of geometry in the present paper.  

In Secs.~\ref{sec:3HeB} and \ref{sec:phononB} of the paper, we calculate the transport coefficients starting from the Boltzmann equations for the evolution of the $^3$He and phonon distribution functions.  The $^3$He-phonon scattering amplitude is well established by measurements from the 1960s~\cite{BM67}-\cite{A69}.  For the high $^3$He concentrations of Refs.~\cite{lamoreaux,rosenbaum}, where $^3$He-$^3$He scattering keeps the $^3$He in equilibribum, we can ignore the energy transfer relative to the momentum transfer, $k$, in the $^3$He-phonon scattering.  However, in general, it is necessary to consider both.  The leading effects appear at order $k^2$, for which the $^3$He Boltzmann equation is effectively a Fokker-Planck equation,  but there are important contributions at order $k^4$, which we describe as recoil corrections (see Sec.~\ref{sec:recoil}).  We find that the solutions of the Fokker-Planck equation for the $^3$He distribution function are exactly Sonine polynomials, familiar from calculations of classical transport coefficients of one component gases.  As we calculate, the diffusion constant varies approximately as $T^{-7}$, is independent of $x_3$, and, including the recoil corrections, has a leading coefficient approximately 70\% as large as that found in Ref.~\cite{lamoreaux}.  The thermal transport is dominated by that of the phonons for the full range of $x_3$; for $x_3\, \ga\, 10^{-6}$  $^3$He-phonon scattering reduces $K_{ph}$ below its low $x_3$ value where it is determined by phonon viscosity.  The $^3$He thermal conductivity is limited by $^3$He-$^3$He scattering for $x_3 \,\ga\, 10^{-6}$ but is, in any case, much smaller than $K_{ph}$.  Finally,  the thermoelectric coefficient, $\chi$, because of its ``off-diagonal'' nature, vanishes to lowest order in $k^2$ due to  the orthogonality of the Sonine polynomials; the $k^4$ corrections are furthermore negligible.


\begin{thebibliography}{20}
\bibitem{nosculpa} In Ref.~\cite{highx} we wrote the effective
$^3$He-phonon scattering rate as $\Gamma/p^2$; we emphasize that
when the energy transfer in $^3$He-phonon collisions is taken into account, the scattering rate is
independent of $p$ and has the value $\Gamma/3m^*T$, which is the value of 
$\Gamma/p^2$ for $p^2$ replaced by its thermal average $3m^*T$.
\bibitem{processes}  Three phonon Landau damping and the Beliaev
process, although they involve phonons alone,  can affect the
rate at which momentum is transferred from phonons to $^3$He. These
processes conserve the total momentum and energy flux of the phonons.
However, the cross section for scattering of phonons by $^3$He atoms is
strongly dependent on the phonon momentum and therefore the total rate
at which momentum is transferred from phonons to $^3$He depends on the
details of the phonon distribution, not just the total momentum of the
phonons.
\end{thebibliography}
\end{document}